\documentclass[preprintnumbers,amsmath,amssymb,preprint]{revtex4}
\usepackage{graphicx}
\usepackage{dcolumn}
\usepackage{bm}

\begin{document}
\title{The C-metric black hole near the IR-brane in the $\mbox{AdS}_4$ space}
\author{I. Zakout}
\affiliation{Department of Physics, Stanford University, Stanford, CA 94305-94309}
\date{\today}

\begin{abstract}
We study the black hole shape using the C-metric solution for a matter trapped near the IR-brane in the $\mbox{AdS}_4$ space. 
In the AdS/CFT duality, the IR-brane is introduced by embedding a 2-brane at the $\mbox{AdS}_4$ radius $z=l$, while the UV-brane is defined by the $\mbox{AdS}_4$ boundary. We find the C-metric solution generates a negative tension IR-brane and a negative thermal energy gas of colliding particles.
We analyze the momentum energy tensor at the $\mbox{AdS}_4$ boundary.
We find that the negative energy black hole solution is entirely unstable even for a small perturbation in the Poincare coordinate space.
However, such a black hole decays very rapidly due to the imaginary part emerges in the ADM mass. This imaginary part appears because of the orbifold constraints of the C-metric solution in the unusual coordinates.
Moreover, this decay rate diverges at the UV-brane. This implies that the black hole evaporates instantaneously otherwise the boundary itself will collapse.
\end{abstract}

\maketitle
\section{Introduction}
The production of black holes in the higher dimensions in the heavy ions collisions physics has received much attention recently\cite{Giddings2,Dimopolous1}. It might be the end of short-range physics. 
In the Randall-Sundrum scenario\cite{Randall1,Randall2}, the 4-dimensional world is a brane embedded in a higher dimensional space. 
The black hole can be produced in the brane and the bulk as well. 
However, the black hole solution for a matter trapped in the 3-brane embedded in AdS$_{5}$ is very important to shed some information about the black hole production in the QCD\cite{Polchinski1}.
It has been argued the supergravity solution with warped metric that approximately AdS in large region corresponds to the high energy scattering in a large-N ${\cal N}$ supersymmetric gauge theory with broken conformal symmetry and partial broken supersymmetry\cite{Maldacena1}.

Unfortunately, the nonlinear gravitational solution of the black hole in $\mbox{AdS}_{5}$ is not available. 
Giddings\cite{Giddings1} has studied a linearized solution of the black hole near the IR-brane, 
where the brane is the end of space. 
He found the black hole shape saturates the Froissart bound cross section. 
When the energy increases, the black hole summit reaches smaller values of $z$ 
further toward the AdS boundary.
They will, however, be completely smeared out over the compact manifold $X$.
At this energy, the approximation of these black holes in a background flat space breaks down. 
Therefore, the nonlinear solution becomes essential to describe the shape of such black hole.
The analysis of a lower dimensional world might give some hints what could be happened in the higher dimensional world. 
Emparan, Horowitz and Myers \cite{Emparan1} used the C-metric solution of two accelerated black holes 
to analyze the exact solution of black hole 
near the 2-brane embedded in $\mbox{AdS}_4$ in the Randall-Sundrum scenario\cite{Randall1,Randall2}.
In their treatment, they produced the 2-brane by removing the space between the brane and the $\mbox{AdS}_4$ boundary. 
The full space is completed by gluing the space between the 2-brane and the $\mbox{AdS}_4$ horizon in both sides of the 2-brane. 

To study the black hole shape near the IR-brane, unlike to Ref.\cite{Emparan1}, 
the infrared 2-brane is embedded by removing the space between the brane and $\mbox{AdS}_4$ horizon and then gluing the space from the IR-brane to the $\mbox{AdS}_4$ boundary in both sides of the 2-brane to complete the space. 
The 2-brane and $\mbox{AdS}_{4}$ boundary become the IR-brane and UV-brane, respectively, in the dual AdS/CFT. 
Although, the present problem has been mentioned shortly in Ref.\cite{Emparan1}, we revisit it more thoroughly in the present work to study the stability of such configuration.
 
The outline is that. In sect. II, we present the analysis of the black hole solution near the IR-brane. At first, we present the orbifold constraints. 
Then we show that the produced thermal gas has a negative thermal energy that diverges at $\mu_{\mbox{max}}$.
Finally, we calculate the asymptotic black hole ADM mass at the boundary. 
The conclusion is presented in sect. III.

\section{The C-metric Black hole near $\mbox{AdS}_4$ IR-brane}

The $\mbox{AdS}_{n+1}$ metric in a large region reads 
\begin{eqnarray}
ds^2\approx \frac{l^2}{z^2}\left[ dz^2+\eta_{\mu\nu}dx^{\mu}dx^{\nu}\right]
+l^2ds^2_{X},
\end{eqnarray}
where $l$ is the $\mbox{AdS}_{n+1}$ radius and $X$ is some appropriate compact manifold. 
At long distance, the smooth geometry given above is truncated in the infrared and an infrared ($n$-1)-brane 
is embedded at this end of space at $z=l$. 
The UV-limit of the geometry is the $\mbox{AdS}_{n+1}$ boundary at $z=0$.
The action in the ($n$+1)-dimensional gravity reads
\begin{eqnarray}
I=M^{n-1}_P\int d^n x dz \sqrt{-\hat{g}}\left(R-\Lambda/M^{n-1}_P\right)
-\int d^n x \sqrt{-g_{\mbox{brane}}} T_{n-1}
+\int d^n x dz{\cal L}_{\mbox{matter}},
\end{eqnarray}
where $M^{n-1}_P=\frac{1}{16\pi G_{n+1}}$.
The variation of this action gives Einstein's equations,
\begin{eqnarray}
\sqrt{-\hat{g}}(R_{MN}-\frac{1}{2}\hat{g}_{MN}R)=&-&\frac{1}{2M^{n-1}_P}\left[\sqrt{-\hat{g}}\hat{g}_{MN}\Lambda
-\sqrt{-g_{\mbox{brane}}}
\hat{g}_{\mu\nu}\delta^{\mu}_M\delta^{\nu}_N \delta(z-l)T_{n-1}\right. \nonumber\\
&-&\left.\sqrt{-\hat{g}}T^{\mbox{matter}}_{MN}\delta^{n-1}(x)\delta(z-l)
\right],
\end{eqnarray}
where $T^{\mbox{matter}}_{MN}$ the momentum energy tensor for the source of matter trapped at the ($n$-1)-brane. 
The Einstein's equations have a delta function $\delta^{n-1}(x)\delta(z-l)$ because of the source of matter localized on the brane and another delta function $\delta(z-l)$ because of the IR-brane.
To deal with these two delta functions, the exact solution needs a very sophisticated Green function tools. 
Nonetheless, it is thought that the problem can be simplified drastically by finding a solution for the black hole 
in the $\mbox{AdS}_{n+1}$ with a source of matter localized 
at $\delta^{(n-1)}(x)\delta(z-l)$ and then embedding the ($n-1$)-brane by somehow. 
Unfortunately, even in this simplified version without any IR-brane, 
it is hard to find a stationary solution in the Poincare coordinates.

There is a solution called the AdS C-metric solution\cite{Plebanski1}, which satisfies the Einstein's equation with a negative cosmological constant in 3+1 space. The C-metric solution reads,
\begin{eqnarray}
ds^{2}=
\frac{l^2}{(y-x)^2} 
\left[-U(y)dt^2+\frac{dy^2}{U(y)}+\frac{dx^2}{G(x)}+G(x)d\phi^2\right],
\end{eqnarray}
where 
\begin{eqnarray}
U(y)=y^2(1-2\mu y),
\end{eqnarray}
and
\begin{eqnarray}
G(x)=1-x^2(1-2\mu x).  
\end{eqnarray}
It is invariant under translations of $t$ and $\phi$. The metric solution is written in the unusual coordinates system $x$ and $y$
and satisfies the Einstein's equation with a negative cosmological constant $R_{AB}=-(3/l^2)g_{AB}$.

However, to understand this solution, we have to analyze the orbifold constraints of these coordinates.
It is seemed that $y$ acts as a radial variable, while $x$ is analogous to a polar coordinate.
The range limits for the unusual variables $x$ and $y$ are crucial to characterize the shape of black hole.
The factor $(y-x)^{-2}$ in front of the metric implies that $y=x$ is infinitely far away from points with $y\ne x$ 
and corresponds to the UV-limit in the dual AdS/CFT.
Furthermore, there is another horizon at $y=0$ which is degenerate and has a zero Hawking temperature. 
This adds an additional constraint given by $0\le y\le \infty$.
The black hole horizon is found at $y_{\mbox{BH}}=\frac{1}{2\mu}$.
The range, $y_{\mbox{BH}}\le y \le\infty$, is the domain of the gravitational collapse. 
Hence, the allowed interval outside the black hole horizon is given by 
$0\le y\le y_{\mbox{BH}}$.

The function $G(x)$ plays an important rule to determine the interval for the $x$-polar orbifold and the periodicity 
for the $\phi$-azimuthal angle.
It should be a positive definite in order the metric has a Lorentz signature. 
For $\mu=0$ case, there is only two roots $x_0=-1$ and $x_1=1$.
The interval for $x$-polar orbifold is defined by $x_0<x<x_1$.
When the 2-brane is embedded at $x=0$, it acts as the IR-boundary limit of the geometry. Hence the space will be defined only for 
$x\ge 0$. Since the space for $x\le 0$ will be discarded, we shall focus the discussion for the positive root $x_1$. 
However, to avoid the canonical singularity at $x=x_1$, the metric can be written as
\begin{eqnarray}
ds^2=\left[\frac{1}{G(x)}dx^2+G(x)d\phi^2\right]_{x=x_1}
\rightarrow d\lambda^2+\lambda^2 d\left(\frac{\phi^2}{a^2_{\mu}}\right),
\end{eqnarray}
where 
\begin{eqnarray}
a_{\mu}=\frac{2}{|G'(x_1)|}.
\end{eqnarray}
Hence periodicity for $\phi$-azimuthal angle becomes
\begin{eqnarray}
\Delta \phi=2\pi a_{\mu} \label{deficit1}.
\end{eqnarray}
This periodicity is $2\pi$ for $\mu=0$.
When $\mu$ becomes finite but small the function $G(x)$ has three roots $x_0$, $x_1$ and $x_2$. 
To keep the Lorentz identity signature, the orbifold interval for the $x$-polar is determined by the roots $x_0$ and $x_1$. The 
root $x_2$ is discarded from the 
discussion since it is not needed any more.
As far $\mu$ increases, the root $x_1$ increases as $1\le x_1\le \sqrt{3}$ for $\mu\le \mu_{\mbox{max}}$ where 
$\mu_{\mbox{max}}=\frac{1}{3\sqrt{3}}$.
However, when $\mu$ exceeds $\mu_{\mbox{max}}$, the root $x_1$ disappears and the function $G(x)$ becomes a positive definite 
for all positive $x$-orbifold $x\ge 0$. 
The periodicity $\Delta \phi$ increases and exceeds $2\pi$ as $\mu$ increases and finally diverges at $\mu_{\mbox{max}}$.
This means that $\mu_{\mbox{max}}$ is the upper limit for the parameter $\mu$.
Furthermore, the point $x=y$ is always located outside the black hole horizon for $\mu\le\mu_{\mbox{max}}$ because of 
$y_{\mbox{BH}}>x_1$.

\begin{figure}
\includegraphics{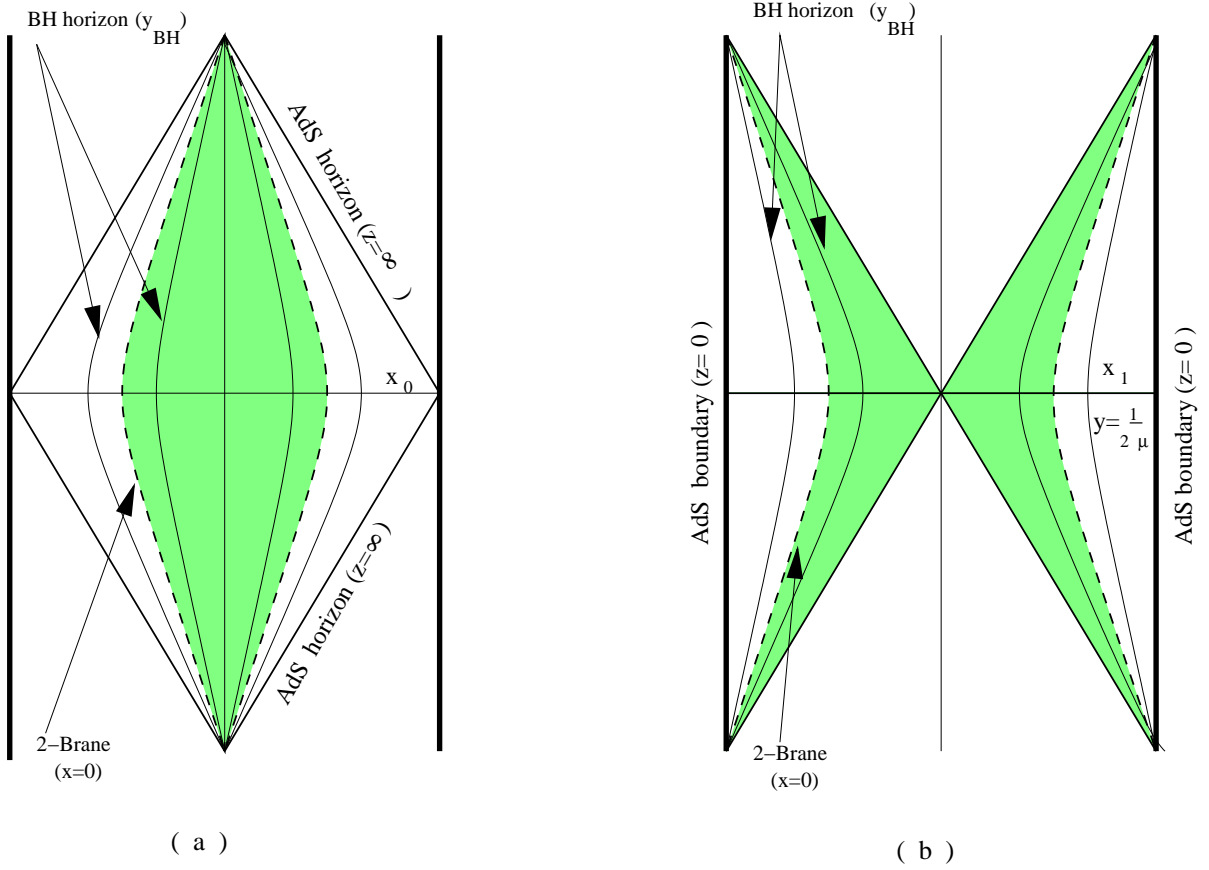}
\caption{\label{fig2}
Pseudo-Penrose diagram shows the cut and glue of the space.
The shaded area is removed from the space.
The space is completed by gluing a mirror copy of the uncut space extended from the brane.
The dashed line represents the thin 2-brane.
The thin solid line represents the black hole horizon.
(a) It covers the space between the 2-brane and the $\mbox{AdS}_4$ 
horizon.
(b) It covers the space between the 2-brane and the $\mbox{AdS}_4$
boundary.}
\end{figure}

To embed the infrared 2-brane at $x=0$, we follow the procedure given in Ref\cite{Emparan1} but instead we cut off the space between
IR-boundary and the $\mbox{AdS}_4$ horizon and then complete the space by gluing onto the surface at $x=0$ 
a mirror copy of the space extended from the IR-boundary to the $\mbox{AdS}_4$ boundary.
The procedure is illustrated by pseudo-Penrose diagram displayed in Fig.\ref{fig2}.
To be more precise, we take a time-like three-surface $\Sigma$ whose 
extrinsic curvature is proportional to its intrinsic metric to be the surface at $x=0$ (or $z=l$) 
and the resulting space-time will be free of conical singularities for the region $x\ge 0$.
Then we take two copies of this side of the space-time and glue them together along $\Sigma$.
The C-metric becomes continuous but not differentiable around $x$.
The extrinsic curvature at $x=0$ is given by\cite{Emparan1},
\begin{eqnarray}
K_{\mu\nu}=\frac{1}{2} n^{\sigma}\partial_{\sigma} \hat{g}_{\mu\nu}|_{x=0}&=&
-\frac{(y-x)\sqrt{G(x)}}{2l} \frac{\partial \hat{g}_{\mu\nu}}{\partial x}|_{x=0}
\nonumber\\
&=&-\frac{1}{l}\hat{g}_{\mu\nu}|_{x=0}.
\end{eqnarray}
The discontinuous in the extrinsic curvature is interpreted as a delta-function $\delta(x)$ (or $\delta(z-l)$) source of stress-energy.
The Poincare coordinate $z$ will be given below.
It is interpreted as a thin relativistic $(n-1)$brane embedded at $x=0$ with 
negative tension $T_{n-1}=-\frac{n-1}{4\pi G_{n+1} l}$ where $G_{n+1}$ is the Newton's constant in 
$n+1$ dimensions. 
This means the brane is expanding out and this is opposed to the physical case where the brane is supposed to has a positive tension. 
The orbifold constraints for the unusual coordinates $x$, $y$ and $\phi$ read
\begin{eqnarray}
0\le &x&\le x_1, \nonumber \\
0\le &y&\le y_{\mbox{BH}}=\frac{1}{2\mu}, \nonumber \\
0\le &\phi& \le 2\pi a_\mu.
\end{eqnarray}
Note that constraint $x\ge 0$ appears because of the 2-brane
is embedded at $x=0$.
There is an additional constraint for a finite matter trapped near the infrared 2-brane
\begin{eqnarray}
\mu\le \frac{1}{3\sqrt{3}}. 
\end{eqnarray}
The black hole horizon never reaches the boundary $y=x$ since $x_1\le x\le y_{\mbox{BH}}$ is forbidden by the orbifold cone constraint. 
The black hole horizon saturates at least $(y_{\mbox{BH}}-x_1)$ away from the $\mbox{AdS}_4$  boundary in the C-metric coordinates.

\begin{figure}
\includegraphics{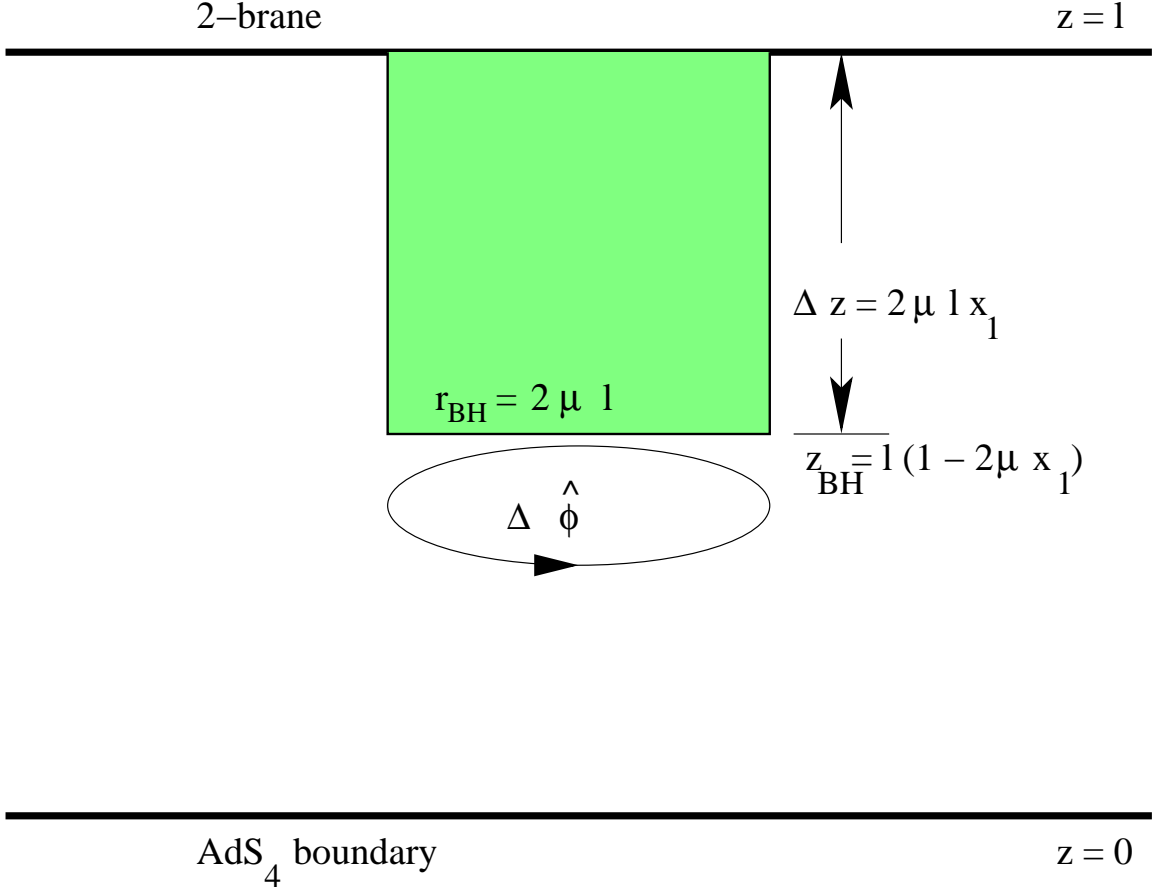}
\caption{\label{fig3}
The black hole solution through the space between the 2-brane 
and the $\mbox{AdS}_4$ boundary.}
\end{figure}

The Poincare transformation is essential to understand the black hole solution in the $\mbox{AdS}_{4}$.
If there is no matter trapped on the IR-brane ($\mu=0$), the function $G(x)$ has only two roots $x_0=-1$ and $x_1=1$. 
Since the region for $x\le 0$ is removed from the space, the allowed interval for the polar coordinate becomes
$0\le x\le 1$. The azimuthal periodicity at $x_1$ is $\Delta \phi=2\pi$.
In this case, the solution in the Poincare coordinates reduces to the $\mbox{AdS}_{4}$ metric 
\begin{eqnarray}
ds^2=\frac{{\it l}^2}{z^2}\left[-d{\hat{t}}^2+dr^2+r^2d\hat{\phi}^2+dz^2\right],
\end{eqnarray}
by using the transformation
\begin{eqnarray}
z&=&{\it l}\frac{(y-x)}{y}, \nonumber \\
r&=&{\it l}\frac{\sqrt{1-x^2}}{y}, \nonumber \\
\hat{\phi}&=&\phi, \nonumber \\
\hat{t}&=&{\it l}t.
\end{eqnarray}
The condition $x=y$ corresponds to the $\mbox{AdS}_{4}$ boundary at $z=0$.
However, The condition $y=0^{+}$ with $-1\le x <0$ corresponds to the usual $\mbox{AdS}_{4}$ horizon at $z=\infty$.
Indeed, we have only the space bounded between the IR-brane at $z=l$ and the UV-brane at $z=0$ 
since the space between the IR-brane and the $\mbox{AdS}_{4}$ horizon is removed.
The curvature singularity at $y=\infty$ is located at the 2-brane $z=l$ and $r=0$ in the Poincare coordinates. 
When a finite amount of matter trapped near the IR-brane (i.e.: $\mu\neq 0$), the solution in the Poincare coordinates becomes
\begin{eqnarray}
ds^2=\frac{l^2}{z^2}\left[-(1-h_{00})d\hat{t}^2 +
ds_{rr}^2+ds^2_{\hat{\phi}\hat{\phi}}+ds_{zz}^2\right],
\end{eqnarray}
where $h_{00}=2\mu y(z,r)$.
The coordinates transformation are modified as
\begin{eqnarray}
z&=&{\it l}\frac{(y-x)}{y}, \nonumber \\
r&=&\frac{l f(x,y)}{y}+2\mu l, \nonumber \\
\hat{\phi}&=&\phi, \nonumber \\
\hat{t}&=&{\it l}t.
\end{eqnarray}
The black hole solution with the C-metric coordinate set transformed
to the Poincare coordinates set is illustrated in Fig.\ref{fig3}.
Note that the azimuthal angle $\phi$ in the C-metric coordinate set is the same
for that $\hat{\phi}$ in the Poincare coordinate set and their periodicity increase and diverges 
at $\mu_{\mbox{max}}$. 
Furthermore the black hole horizon looks like a black cigar solution where its summit is located at $z=l(1-2\mu x_1)$
and reaches it maximum value $z_{max}=l/3$ when $\mu$ reaches its critical value. 
Furthermore, the solution doesn't approach the asymptotic AdS space at the boundary. This means that the deformation of the space because of the negative energy source persists to exist even at the boundary.

The black hole thermodynamics is important to shed some information about the center of mass energy. 
The temperature is calculated as
\begin{eqnarray}
T=\frac{1}{8\pi\mu l}.
\end{eqnarray}
Furthermore, the entropy reads,
\begin{eqnarray}
S=\frac{\cal A}{4G_4},
\end{eqnarray}
where the area is calculated by
\begin{eqnarray}
{\cal A}&=&2l^2\Delta \phi \int_0^{x_1} dx \frac{1}{(y_{\mbox{BH}}-x)^2}.
\end{eqnarray}
The center of mass energy is calculated from the first law of black hole thermodynamics
\begin{eqnarray}
d{\cal E} = T dS.
\end{eqnarray}
It is evaluated as
\begin{eqnarray}
{\cal E}_{\mbox{c.m.}}=G_4M_4=\frac{l}{2}\left(1-\frac{\sqrt{1-2\mu x_1}}{1-3\mu x_1}\right).
\end{eqnarray}
The thermal energy is found negative. It diverges at $\mu=\mu_{\mbox{max}}$ where $3\mu x_1=1$.
Moreover, we have $\sqrt{1-2\mu x_1}>1$ for $\mu\le \mu_{\mbox{max}}$.
The resultant thermodynamical quantities are corresponding to negative energy particles.
However, these results are already found in Ref.\cite{Emparan1} and are expected because of the negative IR-brane tension.
%

When a finite but small amount of matter is trapped near the IR-boundary the solution in the Poincare coordinate is perturbed as follows
\begin{eqnarray}
ds^2&=&\frac{l^2}{z^2}\left[-d\hat{t}^2+dr^2+a_{\mu}^2r^2d\tilde{\phi}^2+dz^2
\right],
\nonumber \\
&=&\frac{l^2}{z^2}\left[h_{ab}dx^adx^b+dz^2\right]=\gamma_{ab}dx^adx^b+\frac{l^2}{z^2}dz^2,
\end{eqnarray}
where $\hat{\phi}=a_\mu\tilde{\phi}$ and $\Delta \tilde\phi=2\pi$.
The stress energy tensor is defined by\cite{Myers2,Balasubramanian1}
\begin{eqnarray}
\tau^{a b}=
\frac{1}{8\pi G_n}\left(\Theta^{a b}-\gamma^{a b }\Theta^c\!_c\right),
\end{eqnarray}
where $a,b,c$ denotes directions parallel to the boundary.
The extrinsic curvature reads
\begin{eqnarray}
\Theta_{ab}=-\gamma_\mu\!^\rho D_{\rho} n_\mu.
\end{eqnarray}
Denote the space time metric as $g_{\mu\nu}$ and $n^{\mu}$ is the outward pointing normal to the boundary normalized with $n^{\mu}n_{\mu}=1$, 
the induced metric on the boundary $\gamma_{\mu\nu}=g_{\mu\nu}-n_{\mu}n_{\nu}$
acts a projection tensor onto the boundary.
The background subtraction procedure yield a finite surface stress tensor\cite{Myers2,Balasubramanian1}
\begin{eqnarray}
\hat{\tau}^{a b}=\tau^{a b}-(\tau^0)^{a b}.
\end{eqnarray}

The conformal transformation can be accounted for by writing the stress tensor expectation values in the field theory as follows
\begin{eqnarray}
\sqrt{-h}h^{ab}<\hat{T}_{bc}>=\lim_{z\rightarrow 0}
\sqrt{-\gamma}\gamma^{ab}\hat{\tau}_{bc},
\end{eqnarray}
where $h_{ab}$ is the background metric of the field theory.
The background metric for the field theory is defined by stripping off the
divergent conformal factor from the boundary
\begin{eqnarray}
h_{ab}=\lim_{z\rightarrow 0} \frac{z^2}{l^2} \gamma_{ab}.
\end{eqnarray}
The total energy in the field theory becomes
\begin{eqnarray}
E_{\mbox{ADM}}=\oint d^{n-2}x \sqrt{-h}<T_{tt}>-\oint d^{n-2}x \sqrt{-h}<T^0_{tt}>.
\end{eqnarray}
When a small amount of matter is introduced, the metric is slightly perturbed by modifying the periodicity of the azimuthal angle. 
The ADM mass at UV-brane becomes
\begin{eqnarray}
E_{\mbox{ADM}}=\lim_{z\rightarrow 0} -\frac{2}{l}\frac{l^3}{z^3}(a_{\mu}-1)\int r dr d\tilde{\phi},
\end{eqnarray}
and
\begin{eqnarray}
E_{\mbox{ADM}}/\mbox{Area}=
\lim_{z\rightarrow 0} -\frac{2}{l}\frac{l^3}{z^3}(a_{\mu}-1).
\end{eqnarray}
The surface energy density diverges even for a very small perturbation of matter. 
This means that present solution is unstable and the space should collapse with any fluctuation.
In C-metric solution, the ADM mass at the boundary $y=x$ is calculated as follows
\begin{eqnarray}
E_{\mbox{ADM}}=\lim_{\zeta\rightarrow 0}
-\frac{2}{l}\frac{l^3}{(\sqrt{2}\zeta)^3}\int d\tilde{\phi}d\xi \sqrt{\frac{2U_0(y)G_0(x)}{U_0(y)+G_0(x)}}
\left[a_{\mu} U(y)/U_0(y) -1\right],
\end{eqnarray}
where
\begin{eqnarray}
y=\frac{\xi+\zeta}{\sqrt{2}}, \nonumber \\
x=\frac{\xi-\zeta}{\sqrt{2}}. 
\end{eqnarray}
If we assume $U(y)/U_0(y)=1$ and $x_1=1$ at the boundary, then the energy becomes
\begin{eqnarray}
E_{\mbox{ADM}}=
\lim_{\zeta\rightarrow 0} -\frac{2}{l}\frac{l^3}{(\sqrt{2}\zeta)^3}
(2/3)2\pi(a_{\mu}-1).
\end{eqnarray}
This quantity diverges at $\zeta=0$.
In the real case, we have $x_1\ge 1$. 
Therefore, the ADM mass has an imaginary part
\begin{eqnarray}
\mbox{Mass}\propto (-\epsilon+i\Gamma_{\epsilon})/\zeta^3.
\end{eqnarray}
The decay rate factor $\Gamma_{\epsilon}$ increases as $\mu$ increases. 
Nonetheless, the decay rate $\Gamma=\frac{1}{\zeta^3}\Gamma_{\epsilon}$ diverges at the boundary $z=0$.

\section{Conclusion}

We have revisited the C-metric solution for a matter trapped near the IR-brane in the $\mbox{AdS}_4$ space bounded by the IR-brane and UV-brane in the AdS/CFT duality.
As noted in Ref\cite{Emparan1}, the C-metric solution generates a negative 2-brane tension and a negative thermal energy gas.
It is claimed that this negative energy gas corresponds to the anti-gravity scenario due to the negative IR-brane tension.
We have demonstrated that the black hole summit never reaches the $\mbox{AdS}_4$ boundary even if its thermal energy diverges at $\mu_{\mbox{max}}$. 
The black hole height tends to saturate at $z_{\mbox{max}}=l/3$. 

We have calculated the asymptotic ADM mass at $\mbox{AdS}_4$ boundary.
We have shown that the C-metric solution implies to entirely unstable configuration in the Poincare coordinates even for a small perturbation. 
The ADM mass diverges at the boundary for the C-metric solution.
Furthermore, the C-metric solution in the unusual coordinates space produces a decay rate factor because of the imaginary part in the ADM mass. 
This imaginary part appears because of the orbifold constraints. 
This decay rate diverges at the $\mbox{AdS}_4$ boundary. 
This means that either the black hole is unstable and evaporates instantaneously or the space configuration is unstable and subsequently the $\mbox{AdS}_4$ boundary collapses.

\acknowledgments 
I am grateful to S. Giddings, S. Shenker and L. Susskind 
for helpful and stimulating discussions. 
I also thanks D. Bak, K. Dasgupta, C. Herdeiro, S. Hirano,  B. 
Kol, M. Sheikh-Jabbari and V. Hubeny for valuable conversations. 
This work is supported by Fulbright FY2002 grant.


\end{document}